\documentclass[reprint,superscriptaddress,amsmath,amssymb,aps,pra,nofootinbib,longbibliography]{revtex4-2}
\usepackage[english]{babel}
\renewcommand{\selectlanguage}[1]{}
\usepackage[utf8]{inputenc}
\usepackage[normalem]{ulem}
\usepackage{physics}
\usepackage{siunitx}
\usepackage{upgreek}
\usepackage[dvipsnames]{xcolor}
\usepackage{graphicx}
\usepackage{dcolumn}
\usepackage[version=4]{mhchem}
\usepackage{bm}
\definecolor{rred}{RGB}{151,23,46} 
\usepackage[colorlinks=true,citecolor=blue,urlcolor=blue,linkcolor=blue]{hyperref}
\usepackage[mathlines]{lineno}
\newcommand{\Rb}{\ce{^{87}Rb} }
\newcommand{\temp}{T}
\renewcommand{\vr}{\mathbf{r}}
\newcommand{\vv}{\mathbf{v}}
\newcommand{\tint}{t_{\rm int}}
\newcommand{\teff}{t_{\rm eff}}

\def\ANU{Department of Quantum Science and Technology, The Australian National University, Canberra, Australian Capital Territory 2601, Australia.}
\def\UVA{Department of Physics, University of Virginia, Charlottesville, Virginia 22904, USA.}
\def\NG{Northrop Grumman Mission Systems, Woodland Hills, California, 91367, USA.}

\begin{document}

\preprint{APS/123-QED}

\title{Ramsey Atom Interferometry with Transverse Optical Modes}

\author{Ryan Husband}
\email[Contact author: ]{ryan.husband@anu.edu.au}
\affiliation{\ANU}

\author{Ryan J. Thomas}
\affiliation{\ANU}%

\author{Yosri Ben-A\"{i}cha}
\affiliation{\ANU}

\author{Rhys H. Eagle}
\affiliation{\ANU}%

\author{Jessica Eastman}
\affiliation{\ANU}%

\author{John E. Debs}
\affiliation{\ANU}%

\author{Patrick J. Everitt}
\affiliation{\ANU}

\author{Michael Larsen}
\affiliation{\NG}

\author{Eric Imhof}
\affiliation{\NG}

\author{Charles A. Sackett}
\affiliation{\UVA}

\author{John D. Close}
\affiliation{\ANU}%

\author{Simon A. Haine}
\affiliation{\ANU}

\author{Samuel Legge}
\affiliation{\ANU}%

\date{\today}

\begin{abstract}
    We experimentally demonstrate atom interferometry using the transverse phase profile of an optical mode. As proof-of-principle, we use the helical phase windings of hypergeometric-Gaussian beams for Ramsey interferometry with ensembles of ballistically-expanding cold \Rb atoms, and we show that the interferometer can measure rotations induced by a motor with a sensitivity that scales linearly with orbital angular momentum and interferometer time. We characterize the thermal decoherence of the interferometer, deriving and experimentally confirming a closed-form expression for the spatially-varying interferometer visibility arising near the singularity of the helical phase winding, motivating the use of condensed atoms in ring-shaped traps.
\end{abstract}

\maketitle

\section{Introduction}
Light pulse atom interferometry (LPAI) enables highly precise \cite{Peters2001HighPrecisionGravity,Altin2013BraggGravimeter,Rosi2014NewtonianGColdAtoms,Hardman2016BECGravMagGrad,Zhang2023BraggLorentz} and accurate \cite{Karcher2018UltracoldSources,Geiger2020AVS,Janvier2022CompactDifferentialGravimeter} measurements of acceleration \cite{Evstifeev2017OnboardGravityGradiometers,Bidel2018MarineGravimetry,Migliaccio2019MOCASS,Bidel2020AirborneGravimetry,Wu2023MarineGravityField}, time \cite{Borde2002Clocks,Hu2017SrClockTransition,Roura2025FreelyFallingClock}, electromagnetic fields \cite{Ekstrom1995PRA}, and rotation \cite{Gustavson2000DualSagnac,Stockton2011GeodeticRotation,Gautier2022SagnacMatterWaves,deCastanet2024RotationRates}, with important results including state-of-the-art measurements of the fine-structure constant \cite{fine_structure_measurement,Morel2020FineStructureConstant}, demonstrations in microgravity \cite{Becker2018SpaceBorneBEC,Mueller2020PlanetaryMissions,AI_in_space}, and use as dark matter detectors \cite{abe_matterwave_2021}. In nearly all LPAIs, the signal is encoded in a laser-imprinted longitudinal phase that is approximately constant across the transverse extent of the atom cloud, yielding a large scale factor and near-uniform coupling over a macroscopic atomic ensemble \cite{KasevichChu1991PRL,off_resonant_raman}. The reliance on the longitudinal phase has left the transverse phase as a systematic effect to be mitigated, and it is primarily treated as a source of bias and dephasing rather than a metrological resource \cite{Schkolnik2015ApplPhysB,Zhou2016PRA,LeGouet,Cronin2009RMP,Bade2018,Gaudout2025}, although controlled transverse phase gradients have also been employed for spatially resolved interferometer readout \cite{phase_shear_AI,Asenbaum2020}. Here, we instead deliberately use a nonlinear transverse phase profile as the primary sensing phase of a Ramsey atom interferometer, enabling sensing with structured optical wavefronts.

Optical wavefronts can be engineered to create potentially useful features such as vortices \cite{circle_beams}, saddle phases \cite{saddle_phase}, high-order mode superpositions \cite{Abramochkin2004GeneralizedGaussian}, and engineered aberrations \cite{Forbes2021NatPhoton,Allen1992PRA}. Access to this form of phase engineering in LPAIs enables wavefront geometries that are sensitive to rotations, transverse forces, and field gradients, since the accumulated phase depends on the specific trajectories of atoms during the interrogation time \cite{LeGouet,multi_axis_AI,point_source_AI,wavefront_mapping_AI}. Such phase control also offers a path for spatially resolved sensing because different regions or distinct transverse modes can encode independent phase information within a single measurement, enabling a full fringe characterisation in a single shot \cite{phase_shear_AI,Yosri_paper}. 

\begin{figure}[ht]
  \centering 
  \includegraphics[scale=1]{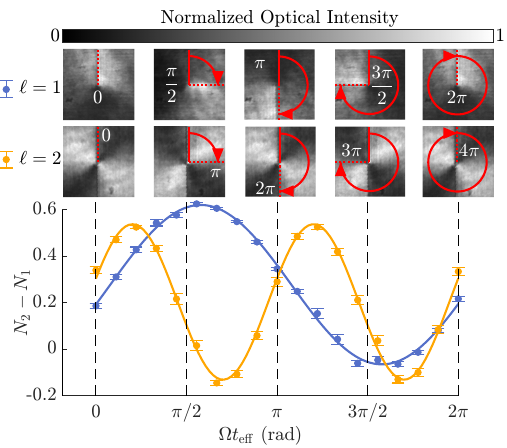}
  \caption{The measured optical interferograms in the top two rows illustrate how the helical phase structure of the optical vortex rotates in the lab frame, with red arrows indicating the angle of rotation and white text indicating the corresponding phase change of the Ramsey fringe, $\phi$, from Eq.~\eqref{eq:phase-shift}. The bottom plot depicts the output of a LPAI using the above optical vortices, where the Ramsey fringes a presented with the normalised atom population difference ($N_2 - N_1$), showing the increase in fringe frequency with $\ell$.}
  \label{fig:thermal_VMG}
\end{figure}

The quantised helical phase gradient of optical vortices, alongside the relative ease with which they can be generated \cite{Wang2018RecentAdvancesOpticalVortex}, makes these beam profiles attractive candidates for a transverse phase LPAI. Several schemes \cite{use_case_4,use_case_3,use_case_2,use_case_1} propose using two optical vortex laser beams to measure a rotation rate, $\Omega$, with a Ramsey interferometer so that the atom interferometer phase is
\begin{equation}
    \phi = \ell\Omega\teff = \ell\Omega\left(\tint + \frac{4\tau}{\pi}\right),
    \label{eq:phase-shift}
\end{equation}
where $\ell$ is the difference in orbital angular momentum (OAM) quantum number between the two vortex beams, $\tint$ is the time the atoms evolve between pulses, and $\tau$ is the length of each $\pi/2$ pulse. An example of the output from this interferometer is demonstrated in Fig.~\ref{fig:thermal_VMG}. Unlike Sagnac-style gyroscopes where the interferometer scale factor explicitly depends on a geometrical area, such as ring laser \cite{chow_ring_1985}, fibre-optic \cite{vali_fiber_1976}, and atom interferometer gyroscopes \cite{gustavson_precision_1997,Gustavson2000DualSagnac,Moan2020WaveguideGyro}, the scale factor in this ``vortex matterwave gyroscope'' depends only on a known integer and an atomically-locked time, making it a fundamentally different type of rotation sensor more akin to NMR gyroscopes \cite{bevan_nuclear_2018}.

Unfortunately, most optical vortices, such as standard Laguerre-Gauss modes \cite{seigman}, feature strong radial intensity gradients due to the phase singularity at the center of the beam, leading to undesirable light-matter interactions for a LPAI relative to using two large Gaussian modes \cite{tophat_interferometry,saywell}. One method for reducing radial intensity variations is to magnify the optical vortex and position the cloud inside the primary ring of the far-field intensity profile. Another is to re-image the beam in the near-field immediately after the phase imprinting plane so that the initial intensity pattern is approximately reproduced. In either case, an incident Gaussian beam results
in a hypergeometric-Gaussian (HyGG) mode \cite{circle_beams,circle_beams2}, which in this near-field regime is known as a Gaussian vortex mode \cite{GV_husband2025}. Both methods significantly reduce the radial intensity gradients while preserving the helical phase required for a transverse phase LPAI.

In this Letter, we experimentally investigate the fundamental properties of the helical phase interferometer on a freely-falling thermal atom cloud using the helical transverse phase of a HyGG mode created using spiral phase plates. We first demonstrate that the interferometer is sensitive to rotations by rotating the far-field phase profile during the interferometer sequence and mapping the resulting interferometer fringes when the atoms are located inside the primary ring of the HyGG mode. We then investigate dephasing of the interferometer due to the thermal motion of the atoms through the transverse phase profile when the atoms are located near the phase singularity of the HyGG mode, which leads to spatially-dependent interferometer contrast. We show that the characteristic length scale over which the contrast drops is described by a simple analytic model which has important implications for the design and use of transverse phase atom interferometers.

\section{Experiment}
\begin{figure}
    \includegraphics[width = \columnwidth]{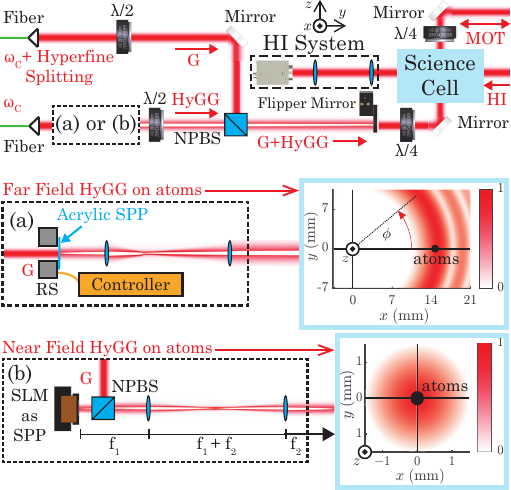}
    \caption{(top) The cold atom and interferometer apparatus as described in the text, where the dashed box labeled (a) or (b) indicates the two different optical vortex generation options which propagate along the vertical $z$ axis. (a) A Gaussian beam becomes a HyGG mode by helical phase modulation using an acrylic spiral phase plate (SPP). The phase plate is mounted to a rotation stage (RS) \cite{NewportRGV100BLS_ProductPage} and set to rotate about a fixed axis by a controller. The HyGG mode is then magnified to cover the atom cloud in the transverse $(x,y)$ plane (to scale). (b) A Gaussian beam retro-reflects from a spatial light modulator (SLM) (Meadowlark Optics model ODPDM512, with a $512\times 512$ pixel array and a pixel pitch of \SI{15}{\micro\meter}) set as a SPP. A $4f$ imaging system ensures that the intensity profile of the Gaussian beam incident on the SLM is positioned onto the transverse $(x,y)$ plane of the atom cloud while maintaining the helical phase of the SPP.}
    \label{fig:experiment_setup}
\end{figure}

We produce cold \Rb{} clouds using the apparatus detailed in \cite{our_BEC_machine,pat_thesis,RT_paper}, where the final evaporation stage in our optical dipole trap produces clouds of $10^6$ atoms with temperatures ranging from \SI{100}{\nano\kelvin} to \SI{10}{\micro\kelvin}. After evaporation, the atoms are prepared in the $\ket{F = 1,m_F = 0}$ state before being released where they are then used in the interferometer. Interferometer fringes are measured from state-resolved imaging of the atoms after the interferometer sequence using the horizontal imaging (HI) system shown in Fig.~\ref{fig:experiment_setup}. A vertical imaging system (not shown) was used to align the optical vortices with the cold atom cloud; however, its shallow depth of field made it unsuitable for data acquisition.

The Raman light is generated from a common \SI{1560}{\nano\meter} fiber laser, with frequency $\omega_\mathrm{C}/2$, which is sent to two home-built Er-doped fiber laser amplifiers. An optical I/Q modulator, locked in the carrier-suppressed single-sideband mode \cite{SSB_lock}, shifts the frequency of the light in one of these paths by half of the hyperfine splitting of \Rb{}. Each fiber path is coupled into free space where the light is converted to \SI{780}{\nano\meter} using second harmonic generating crystals, and whose amplitude and phase is subsequently controlled using independent double pass acousto-optic modulators \cite{double_aom}. Single-mode fibers then clean each beam's mode, after which one beam undergoes helical phase modulation using one of the setups in Fig.~\ref{fig:experiment_setup}(a) or (b) while the other beam remains a Gaussian, so that the difference in orbital angular momentum is set by the orbital phase winding $\ell$ of the HyGG mode's phase plate. As discussed earlier, the near and far-field designs were motivated by the resulting near-uniform intensity profile across the cloud, which is a desirable trait of typical LPAIs \cite{saywell}. We overlap the HyGG beam with the remaining Gaussian beam using a non-polarising beam splitter (NPBS), such that each mode is copropagating through the science cell vertically after the MOT beam flipper mirror moves. 

\section{Measuring Rotations with a Far-Field HyGG Mode}
\begin{figure}[bt]
    \centering
    \includegraphics[width=\columnwidth]{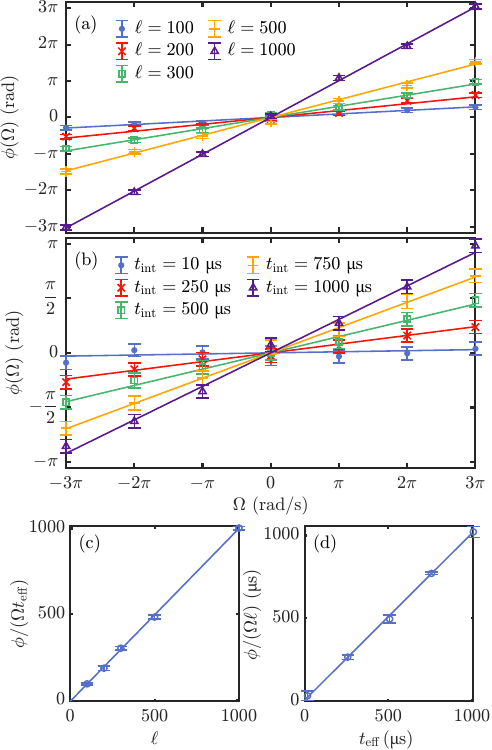}
    \caption{(a--b) Measured interferometer phases $\phi(\Omega)$ due to rotation of the phase plate (markers) with linear fits (solid lines), with the fit offset subtracted for clarity. Measurements for different $\ell$ and a fixed $\tint = \SI{1}{\milli\second}$ are shown in (a) and for different $\tint$ and a fixed $\ell = 300$ in (b). (c) Extracted slope, $\phi/(\Omega\teff)$, from (a) as a function of $\ell$. Solid line is a linear fit to $\phi/\Omega\teff = 1.01(0.01)\ell - 5.6(5.9)$. (d) Extracted slope, $\phi/(\Omega\ell)$, from (b) as a function of $\teff$. Solid line is a linear fit to $\phi/\Omega\ell = 1.01(0.02)\teff + \SI{6.5(16)}{\micro\second}$.}
    \label{fig:motor_measurements}
\end{figure}

We measure physical rotations of the optical fields by rotating the spiral phase plates using the setup shown in Fig.~\ref{fig:experiment_setup}(a). Several phase plates were made using a diamond lathe from acrylic with $\ell$ ranging from $100$--$1000$ \cite{phase_plate}. Rotating the phase plate attached to the stage rotated the output HyGG mode about the phase singularity, and the rotation was applied continuously throughout the measurement time. For this measurement, we use a cloud with a temperature of \SI{1}{\micro\kelvin} and an initial expansion time of \SI{10}{\milli\second} before the interferometer pulses are applied, with an interferometer time of $\tint = \SI{1}{\milli\second}$. For each rotation rate $\Omega$, we measure a fringe comprised of $200$ interferometer shots by randomly modulating the optical phases of a single Raman beam in $1.8^\circ$ increments and counting the atom number in each ground state. By fitting to this fringe, we infer the rotational phase shift induced by the motor, and these measurements were repeated for each phase plate as shown in Fig.~\ref{fig:motor_measurements}a. Additionally, using the $\ell = 300$ phase plate, we measure interferometer fringes with $\tint$ ranging from \SI{10}{\micro\second} to \SI{1}{\milli\second}, seen in Fig.~\ref{fig:motor_measurements}b. For this data set, fringe contrasts are approximately constant as a function of $\ell$, but decrease from 79\% for $\tint = \SI{10}{\micro\second}$ to 45\% for $\tint = \SI{1}{\milli\second}$. As expected, the phase shifts as a function of $\Omega$ are highly linear. The uncertainties on each phase measurement combine the 1-sigma sinusoidal fit uncertainty with a systematic contribution from a bias phase drift, measured to be \SI{0.13}{\radian} over the \SI{2.2}{\hour} timescale between successive fringe measurements. In the most sensitive case, the $\ell = 1000$ fringe at $\tint =$ \SI{1}{\milli\second} exhibited a single shot phase noise of \SI{0.47}{\radian}, extracted from the residual variance of the fringe fit. Combined with the scale factor measured for this configuration, this corresponds to a rotation resolution of $\Omega = 0.47$ rad/s per shot. 

Scale factor accuracy is an important performance metric for any atom interferometer. In longitudinal phase interferometers, the scale factor accuracy is primarily set by the accuracy to which the laser wavevector is known, and this is normally locked to an atomic transition although small corrections exist due to the finite beam size \cite{cervantes_selection_2021}. In our helical phase interferometer, the scale factor accuracy is primarily set by the purity of the optical vortex mode, with ideal interferometers having modes associated with a single OAM value of $\ell$. Mode impurities associated with additional OAM values modify $\ell$ based on the relative power in those modes. We extract the interferometer scale factor by linearly fitting each data set shown in both Figs.~\ref{fig:motor_measurements}a and \ref{fig:motor_measurements}b to measure the slope of the lines of best fit, and we then fit straight lines to $\phi/\Omega$ for either fixed $\ell$ or fixed $\teff$ to the slope measurements as shown in Figs.~\ref{fig:motor_measurements}c and \ref{fig:motor_measurements}d, respectively. Both measurements show excellent agreement with Eq.~\eqref{eq:phase-shift}, confirming that mode impurities have a negligible impact on the scale factor accuracy and that the interferometer response is linear in time, as expected for a pure rotation rate measurement.

\section{Measuring Thermal Dephasing with a Near-Field HyGG Mode}
\begin{figure}[bt]
    \centering  
    \includegraphics[width=\columnwidth]{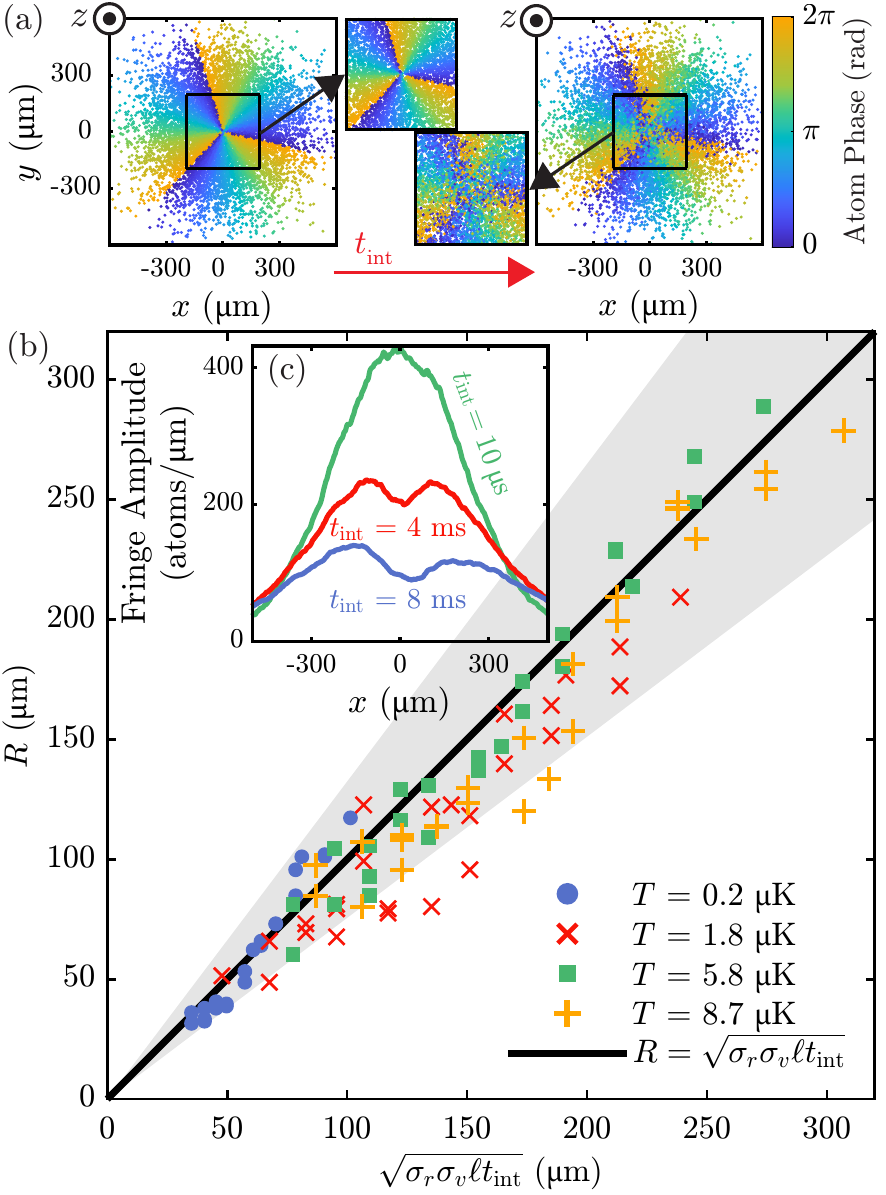}
    \caption{(a) Schematic illustration of the effect of thermal motion of the atoms on the interferometer contrast across the cartesian space $(x,y)$. Atoms are initially imprinted with a helical phase (shown as variation in colour), and the atoms in the center of the cloud move more than those at the edges, obscuring the interferometer phase and leading to contrast reduction. (b) The dephasing radius $R$ measured as a function of $\sqrt{\sigma_r\sigma_v\ell \tint}$ over all $\ell$, $\tint$, and $\temp$. The markers are the measured data, and the black solid line is Eq.~\eqref{eq:dephasing-radius} with no free parameters and with the gray shaded area showing the one standard deviation model uncertainty based on uncertainties in $\temp$ and $\sigma_r$. (c) An example of the central dip in fringe amplitude due to the phase singularity for the $\ell = 3$ and $T = \SI{5.8}{\micro\kelvin}$ dataset, as a function of the transverse cloud axis $x$, from (a). The slight asymmetry for $\tint = \SI{8}{\milli\second}$ is due to a small misalignment between the optical mode and the atomic cloud.}
    \label{fig:thermal_data}
\end{figure}

Transverse phase variations in LPAIs can couple to the thermal motion of the atomic ensemble, causing different atoms to accumulate different phases and reducing the measured contrast, although controlled phase-shear readout can mitigate this loss when the phase remains correlated with the detected atomic position \cite{Asenbaum2020}.  In our particular implementation with a helical transverse phase, an atom located at position $\mathbf{r}$ with velocity $\mathbf{v}$ acquires a phase
\begin{equation}
\phi(\mathbf{r},\mathbf{v})=\ell t_{\mathrm{eff}}\left[\Omega-\dot{\theta}(\mathbf{r},\mathbf{v})\right],
\end{equation}
where $\dot{\theta}=(\mathbf{v}\cdot\hat{\mathbf{y}}\cos\theta-\mathbf{v}\cdot\hat{\mathbf{x}}\sin\theta)/r$ is the angular velocity of the atom, $\theta$ is its polar angle, and $r$ is its radial separation from the phase singularity. Since $\dot{\theta}$ scales inversely with $r$ for a given transverse velocity, atoms that are closer to the helical phase singularity have larger angular velocities which lead to larger phase shifts, as seen in Fig.~\ref{fig:thermal_data}a, so we expect thermal dephasing to be more pronounced for these atoms compared to those at larger separations. 

We demonstrate this spatially-dependent thermal dephasing using the HyGG architecture shown in Fig.~\ref{fig:experiment_setup}b, where we center the re-imaged near-field HyGG mode on the atomic distribution. As highlighted in \cite{GV_husband2025}, the finite aperture of the optical system filters high spatial‑frequency components of the mode, leading to the formation of a vortex core whose size increases with $\ell$. The acrylic phase plates used in the motor measurements produced vortex cores larger than the atom cloud at the imaging plane, so instead we use phase plates with low $\ell = 1-5$ from the SLM. The resulting vortex core sizes are below the \SI{5.5}{\micro\meter} pixel-limited resolution of our imaging system and are also negligible compared to the atomic cloud widths of order \SI{100}{\micro\meter}.

Atom clouds were released from the optical trap and allowed to expand ballistically for \SI{10}{\milli\second} prior to the first $\pi/2$ pulse, ensuring that the cloud size at the start of the interferometer sequence is dominated by the thermal velocity distribution rather than the initial trap geometry. We map a fringe for each $\tint$ using the same technique as for the rotation measurements, and we image the atoms \SI{100}{\micro\second} after the final $\pi/2$ pulse. Note that from Fig.~\ref{fig:experiment_setup} each absorption image of the atom cloud is a projection along the $y$-axis, into the $(x,z)$ plane.  To see the interferometer dephasing near the helical phase singularity (along $z$), we integrate the measured atomic densities along the vertical axis $z$, to produce one-dimensional atom densities along the horizontal direction $x$. By mapping out an interferometer fringe, this image analysis yields an interferometer fringe at each position of $x$. By fitting to each fringe profile, we find the amplitude as a function of $x$ to consist of an overall Gaussian shape from the distribution of the atom cloud alongside a pronounced dip in amplitude towards the center of the cloud, as seen in Fig.~\ref{fig:thermal_data}c, coinciding with the optical phase singularity at $r = 0$. We quantify the size of this dephasing region by half of the distance, $R$, between the peaks in the one-dimensional fringe amplitude. Assuming a Maxwell-Boltzmann velocity distribution for the atomic ensemble, $R$ obeys the analytical expression (derived in the Appendix)
\begin{equation}
    R \approx \sqrt{\sigma_r\sigma_v \ell \tint},
    \label{eq:dephasing-radius}
\end{equation}
where $\sigma_r$ is the cloud width at the trap release point inferred from time-of-flight measurements assuming purely ballistic expansion, and $\sigma_v = \sqrt{k_\mathrm{B}\temp/m}$ is the velocity width, with $m$ as the mass of a \Rb{} atom and $\temp$ as the ensemble temperature. In Fig.~\ref{fig:thermal_data} we plot the measured radius $R$ as a function of $\sqrt{\sigma_r\sigma_v \ell \tint}$ for all $\ell$, $\tint$, and $\temp$, demonstrating excellent agreement with Eq.~\eqref{eq:dephasing-radius}.

\section{Discussion and Outlook}
Our choices of optical vortex architecture illustrate how the transverse optical phase can be used for atom interferometry and, in particular, inertial sensing. We emphasize that the value of this architecture is not primarily in competing with the sensitivity of large-area Sagnac atom gyroscopes \cite{Barrett2019PRL,Savoie2018InterleavedGyro}, which benefit from decades of optimization of a geometrically distinct approach, but in its scale factor, set by an integer OAM difference and an atomically-locked time rather than an enclosed area (Eq.~\eqref{eq:phase-shift}). This makes it, to our knowledge, one of two demonstrated primary rotation-sensing architectures alongside NMR gyroscopes \cite{article}, whose scale factor requires no external calibration, suggesting a natural role alongside a classical gyroscope, correcting long-term drift while the classical sensor supplies bandwidth. A general drawback of modes with structured transverse phases is the resulting intensity variations that typically reduce interferometer visibility. Optical vortex modes exemplify this effect, as the phase singularity always corresponds to an intensity null. As demonstrated here, this issue can be mitigated by re-imaging the phase imprinting plane onto the atoms, with the caveat that limited optical apertures will prevent the intensity variations from being suppressed entirely \cite{GV_husband2025}. Trade-offs on the optical aperture can then be made based on ensuring that intensity variations associated with the singularity occur at spatial scales smaller than the dephasing radius, since making the intensity variation region any smaller confers limited additional benefit to the interferometer performance.

A complementary approach is to tailor the atomic density distribution to avoid regions where these contrast-limiting effects are significant. For example, in our demonstration of Fig.~\ref{fig:motor_measurements} we simply offset the atomic distribution from the center of the mode, so that the atoms avoided both the null-intensity region and the phase diffusion region for the interferometer times we considered. An equivalent approach would be to arrange the atoms in a toroid around the optical mode using a modified optical dipole trap: this approach has the advantage that the trap and Raman beams could be generated using the same optics, avoiding geometrical mismatches between the atomic distribution and the optical mode. More complex transverse phase profiles would require commensurately more complex atomic distributions, but these can be generated using painted potentials \cite{henderson_experimental_2009,Bell_2016} or digital micromirror devices \cite{gauthier_direct_2016,woffinden_paper}. 

Ultimately, the sensitivity of transverse phase LPAIs will be limited by the temperature of the atomic source, especially for Ramsey interferometers using ballistically-expanding ensembles, since phase diffusion effects are strongly controlled by temperature. Bose-condensed sources significantly alleviate this problem in transverse phase interferometers, as they do in conventional LPAI approaches \cite{Hardman2016BECGravMagGrad,stuart_paper}, with the disadvantage that they require more complex apparatuses and longer sample preparation times. This overhead can be substantially reduced using atom-chip BEC sources \cite{BEC_chips}, which, combined with in-trap operation, would remove the need for the long free-fall baseline and associated large vacuum systems that conventional cold-atom Sagnac gyroscopes require, pointing toward a compact, primary rotation sensor rather than a laboratory-scale one. For rotation sensing in particular, condensed atoms in a toroidal trap can be prepared as long-lived counter-rotating currents \cite{use_case_1,use_case_2,use_case_3,use_case_4}, with long interferometer times limited not by thermal phase diffusion but instead by atomic interactions and the dynamics of the trap \cite{use_case_2}. More broadly, different transverse phase LPAIs may see similar benefits when used with trapped, condensed atomic ensembles. Our demonstration represents a small part of the broader potential offered by transverse optical modes where the ability to design arbitrary transverse phase profiles opens a wide parameter space for new interferometer architectures and applications beyond those accessible with longitudinal phase interferometers.

\section*{Acknowledgments}
This research was supported by the Australian government Department of Industry, Science, and Resources via the Australia-India Strategic Research Fund (Grant No. AIRXIV000025), and Australian Research Council Grants No. FT210100809 and No. LP190100621.

\appendix*
\section{Derivation of the Near-Field HyGG Dephasing Radius}\label{appendix}

\begin{figure}[ht]
  \centering
  \includegraphics[scale=1]{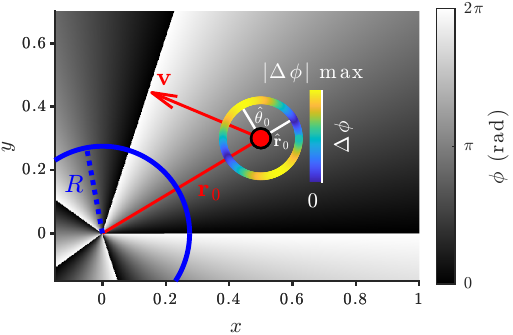}
  \caption{A single thermal atom (red circle) is pictured at radius $\mathbf r$ moving at velocity $\mathbf v$ across the helical phase front, $\phi$, of a GV mode ($\ell = 5$) in the VMG. During $t_{\mathrm{int}}$, the atom will exhibit an interference phase $\Delta\phi$, determined by the immediate direction of motion (the colored ring). The dephasing region radius $R$ is shown in blue.}
  \label{fig:dephasing_model}
\end{figure}

We consider the effect of the thermal velocity of the atoms on the spatially dependent contrast of the interferometer, as seen in Fig. \ref{fig:dephasing_model}. For this calculation, we ignore the spatial variation of the intensities of the Raman beams so that their only contribution is the azimuthal phase $\phi(\mathbf r) = \ell\theta$ for polar angle $\theta$. We assume that the center of the phase winding and the center of the atomic distribution are at the same location in space, and we restrict the model to 2D space since the longitudinal direction of the optical modes plays a negligible role in the interferometer contrast. The Raman beams imprint a phase $\phi(\vr_1)$ onto the atomic cloud at $t=t_1$ using a $\pi/2$ pulse, where $\vr_1 = \vr(t_1)$ is the position of an atom at time $t_1$, and similarly imprint $\phi(\vr_2)$ at $t_2=t_1+t_{\mathrm{int}}$ where $\vr_2 = \vr(t_2)$.

For an individual atom, the phase accumulated during $t_{\mathrm{int}}$ is
\begin{equation}
\begin{split}
\Delta\phi &= \phi(\vr_2) - \phi(\vr_1)\\
&= \ell\,\big[\theta(t_2) - \theta(t_1)\big].
\end{split}
\label{eq:phase-diff-theta}
\end{equation}
Defining the polar coordinates through $x = r\cos\theta$ and $y = r\sin\theta$, so that $r=\sqrt{x^2+y^2}$ and $\theta=\tan^{-1}(y/x)$, the corresponding velocities are
\begin{align}
\dot r &= \dot x\cos\theta + \dot y\sin\theta, \\
\dot\theta &= \frac{\dot y\cos\theta - \dot x\sin\theta}{r},
\end{align}
and their inverses are
\begin{align}
\dot x &= \dot r\cos\theta - r\,\dot\theta\sin\theta, \\
\dot y &= \dot r\sin\theta + r\,\dot\theta\cos\theta.
\end{align}
The change in angle between any two points in time is then
\begin{align}
\Delta\theta &= \theta(t_2) - \theta(t_1) = \tan^{-1}\!\left(\frac{r_1\,\dot\theta\,t_{\mathrm{int}}}{r_2}\right),
\end{align}
for $r_2 = |\vr_2|$ and which can be approximated as
\begin{equation}
\Delta\theta \approx \dot\theta\,t_{\mathrm{int}} - \frac{\dot r\,\dot\theta}{r_2}\,t_{\mathrm{int}}^2,\label{eq:dtheta-approx}
\end{equation}
when $r_2 \gg (\dot r,\, r_2\dot\theta)t_{\mathrm{int}}$. Therefore, for each individual atom the population difference between the two states is
\begin{equation}
\Delta P(r_2,\dot r,\dot\theta) = \cos\Delta\phi = \cos\big[\ell\Delta\theta(r_2,\dot r,\dot\theta)\big],
\end{equation}
which depends not only on the radial distance from the center of the cloud but also on the angular and radial velocities. A useful way to represent this is to write it in terms of a complex amplitude
\begin{equation}
\Delta P = \Re\!\left\{ e^{i\Delta\phi} \right\},
\end{equation}
and then the contrast for a single atom can be expressed as
\begin{equation}
C = \left| e^{i\Delta\phi} \right|.
\end{equation}

An atomic ensemble is characterized by its phase-space density function (PSD), $f(\mathbf r, \mathbf v, t)$. Just after the atoms are released from the trap at time $t=0$, the PSD is given by a Maxwell--Boltzmann distribution
\begin{equation}
 f(\vr, \vv, t=0) = \alpha\exp\!\left[-\frac{\vr^2}{2\sigma_{r}^2} - \frac{\vv^2}{2\sigma_{v}^2}\right],
\end{equation}
for spatial widths $\sigma_{r}$ and $\sigma_{v}$ and $\alpha = 1/\left((2\pi)^2\,\sigma_{r}^2\,\sigma_{v}^2\right)$. As the cloud ballistically expands, correlations develop between $\vr$ and $\vv$ through $\vr(t)= \vr + \vv t$. The distribution at $t_2$ is then
\begin{equation}
 f(\mathbf r_2, \mathbf v, t_2) = \alpha \exp\!\left[-\frac{(\mathbf r_2 - \mathbf v t_2)^2}{2\sigma_{r}^2} - \frac{v^2}{2\sigma_{v}^2}\right],
\end{equation}
and the complex fringe amplitude obtained by counting atoms within an infinitesimal area $d^2\vr_2$ around $\mathbf r_2$ is
\begin{equation}
    dA(\mathbf r_2,t_2) = d^2 \vr_2\, \int \! d^2 \vv\; f(\vr_2, \vv, t_2)\, \exp\!\big[ i\ell\Delta\theta(r_2, \dot r, \dot\theta) \big].
\end{equation}
Since the phase shift is best expressed in polar coordinates, we can rewrite the PSD at $t_2$ as
\begin{equation}
\begin{split}
f(r_2, \theta_2, \dot r, \dot\theta, t_2) &= \alpha \exp\Bigg[ -\frac{r_2^2}{2\sigma_{r}^2(t_2)} - \frac{r_2^2\dot\theta^2\,\sigma_{r}^2(t_2)}{2\sigma_{r}^2\sigma_{v}^2} \\
&- \frac{\sigma_{r}^2(t_2)}{2\sigma_{r}^2\sigma_{v}^2}\left( \dot r - \frac{r_2\sigma_{v}^2 t_2}{\sigma_{r}^2(t_2)} \right)^2 \Bigg],
\end{split}
\end{equation}
where $\sigma_{r}^2(t_2) = \sigma_{r}^2 + \sigma_{v}^2 t_2^2$ is the position-space width at time $t_2$. We must also transform our volume element from $dx_2\,dy_2\,dv_x\,dv_y \to r_2^2\,dr_2\,d\theta\,d\dot r\,d\dot\theta$, where one factor of $r_2$ is from the position elements and one is from the velocity elements. Given that the PSD is independent of $\theta_2$, so that only the variation in $r_2$ is relevant, we compute the fringe amplitude $A(r_2,t_2)$ in the region $r_2$ to $r_2 + dr_2$ as
\begin{equation}
\begin{split}
 A(r_2,t_2) &= r_2^2\,dr_2\, \alpha \int_0^{2\pi} d\theta_2 \iint\limits_{-\infty}^{\infty} \exp\Bigg[ -\frac{r_2^2}{2\sigma_{r}^2(t_2)} \\
 &- \frac{r_2^2 \dot\theta^2\,\sigma_{r}^2(t_2)}{2\sigma_{r}^2\sigma_{v}^2}-\frac{\sigma_{r}^2(t_2)}{2\sigma_{r}^2\sigma_{v}^2}\left( \dot r - \frac{r_2\sigma_{v}^2 t_2}{\sigma_{r}^2(t_2)} \right)^2 \\
 &+ i\ell\,\Delta\theta \Bigg] d\dot r\, d\dot\theta \\
 &\approx 2\pi r_2^2\,dr_2\, \alpha\iint\limits_{-\infty}^{\infty}\exp\Bigg[ -\frac{r_2^2}{2\sigma_{r}^2(t_2)}- \frac{r_2^2 \dot\theta^2\,\sigma_{r}^2(t_2)}{2\sigma_{r}^2\sigma_{v}^2} \\
 &- \frac{\sigma_{r}^2(t_2)}{2\sigma_{r}^2\sigma_{v}^2}\left( \dot r - \frac{r_2\sigma_{v}^2 t_2}{\sigma_{r}^2(t_2)} \right)^2 + i\ell\,\dot\theta\,t_{\mathrm{int}} \Bigg] d\dot r\, d\dot\theta,
 \end{split}
\end{equation}
where the approximation is that we drop the $\dot r\,\dot\theta\,t_{\mathrm{int}}^2/r_2$ term in Eq.~\eqref{eq:dtheta-approx}. Evaluating the double integral yields 
\begin{equation}\label{eq:A}
\begin{split}
A(r_2,t_2) = 2\pi\frac{r_2\,dr_2}{\sigma_{r}^2(t_2)} \exp\Bigg[-\frac{r_2^2}{2\sigma_{r}^2(t_2)}- \frac{\sigma_{r}^2\sigma_{v}^2}{2\sigma_{r}^2(t_2)}\,\frac{\ell^2 t_{\mathrm{int}}^2}{r_2^2} \Bigg].
\end{split}
\end{equation}
The contrast function is then obtained by normalizing to the number of atoms in the particular area element, which gives
\begin{equation}\label{eq:contrast}
    C(r_2) = \exp\left[-\frac{\sigma_{r}^2\sigma_{v}^2}{2\sigma_{r}^2(t_2)}\,\frac{\ell^2 t_{\mathrm{int}}^2}{r_2^2} \right].
\end{equation}
Since the contrast function is strictly monotonic away from the origin, it presents no discernible features that can serve to define a region in which dephasing due to the phase singularity is dominant. We therefore turn to the amplitude function, which possesses more readily identifiable structure, including a radial maximum, that is better suited to practical measurements. Experimentally, the cloud is imaged in the plane of the orbital phase, meaning the visibility dip characteristic of the phase singularity must be analyzed as a projection of the 2D atomic density onto the $(x,z)$ plane. Integrating Eq.~\eqref{eq:A} over all $y$ at fixed $x$ yields the projected amplitude
\begin{equation}\label{eq:A_exact}
\begin{split}
A(x,t_2) =\frac{1}{\pi\sigma_{r}^2(t_2)}&\int_0^\infty\exp\!\Bigg[-\frac{x^2+y^2}{2\sigma_{r}^2(t_2)} \\
&-\frac{\sigma_{r}^2\sigma_{v}^2\ell^2t_{\mathrm{int}}^2}{2\sigma_{r}^2(t_2)\,(x^2+y^2)}\Bigg]dy. \\
\end{split}
\end{equation}
which does not have a closed form solution. However, the cloud density is centered on the phase singularity and is radially symmetric, meaning the dominant contribution to the visibility across $x$ occurs on the cross section where $y=0$. Taking advantage of this to approximate a solution to Eq.~\eqref{eq:A_exact}, we first differentiate with respect to $x$ as we would to identify the position of the maximum of $A$ on the $x$-axis, 
\begin{equation}
\begin{split}
\frac{dA_x}{dx}&=\frac{x}{2\pi\sigma_{r}^4(t_2)}\int_{-\infty}^{\infty}\exp\!\Bigg[-\frac{x^2+y^2}{2\sigma_{r}^2(t_2)} \\
&-\frac{\sigma_{r}^2\sigma_{v}^2\ell^2t_{\mathrm{int}}^2}{2\sigma_{r}^2(t_2)(x^2+y^2)}\Bigg]\Bigg[\frac{\sigma_{r}^2\sigma_{v}^2\ell^2t_{\mathrm{int}}^2}{(x^2+y^2)^2} - 1\Bigg]dy.
\end{split}
\end{equation}
By equating the above to zero, we find that the stationary points satisfy
\begin{equation}
\begin{split}
    x &= 0,\quad\textrm{or} \\
    \int_{-\infty}^{\infty}\exp\Bigg(-\frac{x^2+y^2}{2\sigma_{r}^2(t_2)}&-\frac{\sigma_{r}^2\sigma_{v}^2\ell^2t_{\mathrm{int}}^2}{2\sigma_{r}^2(t_2)(x^2+y^2)}\Bigg) \\
    & \times\left[\frac{\sigma_{r}^2\sigma_{v}^2\ell^2t_{\mathrm{int}}^2}{(x^2+y^2)^2}-1\right]dy=0. \\
\end{split}
\end{equation}
Assuming that only atoms near $y = 0$ contribute, the maximum in the fringe amplitude occurs when
\begin{equation}
\begin{aligned}
\frac{\sigma_{r}^2\sigma_{v}^2\ell^2t_{\mathrm{int}}^2}{x^4}-1 = 0\Rightarrow R^2 = \sigma_{r}\sigma_{v}\ell t_{\mathrm{int}},
\end{aligned}
\end{equation}
which is the result cited in Eq.~\eqref{eq:dephasing-radius}.

\onecolumngrid
\vspace{1em}
\begin{center}
\rule{0.5\textwidth}{0.4pt}
\end{center}
\vspace{0.5em}
\twocolumngrid

\renewcommand{\bibsection}{}
\hypersetup{urlcolor=blue}
\nocite{apsrev42Control}
\bibliography{biblio}

\end{document}